\documentclass[aps,prl,superscriptaddress,english,10pt,noshowpacs,twocolumn]{revtex4}
\usepackage[latin1]{inputenc}
\usepackage{graphicx}
\usepackage{float}
\usepackage{latexsym}
\usepackage{graphicx}
\usepackage{amssymb}
\usepackage{amsmath}
\usepackage{amsfonts}
\usepackage{hyperref}
\usepackage{cool}
\usepackage{dcolumn}
\usepackage{textcomp}
\usepackage{color}
\usepackage{xfrac}
\usepackage{changepage}
\DeclareMathAlphabet\mathbfcal{OMS}{cmsy}{b}{n}
\usepackage{xfrac}
\usepackage{nicefrac}
\usepackage{empheq,fancybox}  
\usepackage{amsmath}	
\usepackage{bm}		    
\usepackage{bbm}
\usepackage{amssymb}	
\usepackage{makeidx}	
\usepackage{color} 		
\usepackage{graphicx} 	
\usepackage{enumitem}
\usepackage{multirow}	
\usepackage{tensor} 	
\usepackage{hyperref} 	
\usepackage{slashed}
\usepackage[all]{hypcap}
\usepackage[title]{appendix}
\usepackage{dsfont}
\usepackage{notoccite}
\hypersetup{
    colorlinks=true,	
    linkcolor=red,		
    citecolor=blue,		
    filecolor=magenta,	
    urlcolor=black	    
}

\newcommand{\com}{\ , \ }

\newcommand{\half}{{\textstyle\frac{1}{2}}}

\newcommand{\Mpl}{M_{\rm Pl}}

\usepackage{accents}

\newcommand{\Half}{\nicefrac{1}{2}}

\usepackage{slashed}

\begin{document}

\title{The Gravitino Swampland Conjecture}

\author{Edward W.\ Kolb}
\affiliation{Kavli Institute for Cosmological Physics and Enrico Fermi Institute, The University of Chicago, Chicago, IL 60637}

\author{Andrew J.\ Long}
\affiliation{Department of Physics and Astronomy, Rice University, Houston, TX 77005}

\author{Evan McDonough}
\affiliation{Kavli Institute for Cosmological Physics and Enrico Fermi Institute, The University of Chicago, Chicago, IL 60637}

\begin{abstract}
We extend the swampland from effective field theories (EFTs) inconsistent with quantum gravity to EFTs inconsistent with quantum {\it supergravity}. This enlarges the swampland to include EFTs that become inconsistent when the gravitino is quantized. We propose the {\it Gravitino Swampland Conjecture}:  the gravitino sound speed must be non-vanishing in all EFTs that are low energy limits of quantum supergravity. This seemingly simple statement has important consequences for both theories and observations.   The conjecture is consistent with and supported by the KKLT and LVS scenarios for moduli stabilization in string theory. 
\end{abstract}

\maketitle

{\bf Introduction.}  The swampland program \cite{Vafa:2005ui,Ooguri:2006in} seeks to circumscribe the set of four-dimensional EFTs that are a low energy limit of quantum gravity, e.g, the landscape of  superstring theory vacua \cite{Bousso:2000xa,Susskind:2003kw}, and distinguish these theories from those that are not, thereby enhancing the predictive power of quantum gravity, and in particular, superstring theory. This is done by enumerating criteria that an EFT must satisfy in order to be in the landscape, rather than be relegated to the ``swampland.''  In this work we extend the swampland program from quantum gravity to quantum supergravity, and consider the criteria that emerge from quantization of the fermionic superpartner of the graviton, namely, the gravitino.

The observed cosmological constant implies that local supersymmetry must be non-linearly realized in our universe. The supergravity theory which describes this is de Sitter (dS) Supergravity \cite{Bergshoeff:2015tra,Hasegawa:2015bza, Kallosh:2015sea, Kallosh:2015tea, Schillo:2015ssx,Freedman:2017obq}. This theory was shown in \cite{Kallosh:2014wsa, Bergshoeff:2015jxa, Vercnocke:2016fbt, GarciadelMoral:2017vnz,Cribiori:2019hod,Dasgupta:2016prs} to be the low energy effective field theory of the KKLT proposal for de Sitter vacua in string theory  \cite{Kachru:2003aw}. In the unitary gauge, pure dS supergravity describes a massive gravitino and the graviton. In this sense, the massive gravitino is intrinsic to the supergravity description of our universe. In cases where the gravitino mass is constant, such as in simple models of cosmic acceleration in dS supergravity, the gravitino is well described by the Rarita-Schwinger model  \cite{Rarita:1941mf}.

The canonical kinetic term for the gravitino along with the constraint equations  lead to a modified dispersion relation for the helicity-1/2 gravitino \cite{Giudice:1999yt,Giudice:1999am, Kallosh:1999jj,Kahn:2015mla}, which corresponds to the goldstino of supersymmetry breaking. The sound speed $c_s$ determines the energy associated with a spatial momentum $k$; $c_s k$, with $c_s\leq 1$ in units of the speed of light. In \cite{Kolb:2021xfn}, it was shown that for certain supergravity models and certain parameter regimes, the sound speed can {\it vanish} in the early universe during the cosmological evolution following cosmic inflation ~\cite{Starobinsky:1980te,Sato:1980yn,Guth:1980zm, Linde:1981mu,Albrecht:1982wi,Linde:1983gd}. In supergravity models that reduce to the massive Rarita-Schwinger model \cite{Rarita:1941mf}, a sufficient condition for the sound speed to vanish is $m \lesssim H_e$, where $H_e$ is the Hubble constant at the end of inflation.  

The vanishing sound speed, $c_s^2 =0$, implies that the energy per field excitation is independent of momentum $k$, allowing for the production of particles with arbitrarily high momentum. As argued in  \cite{Kolb:2021xfn}, this implies a breakdown of the EFT. The breakdown of the theory when the gravitino is quantized suggests an extension of the swampland program to include EFTs which become inconsistent when supergravity is quantized. To this end, we propose the Gravitino Swampland Conjecture (GSC):
\begin{quote}  
{\it In all 4-d effective field theories that are low-energy limits of quantum gravity, at all points in moduli space and for all initial conditions, the sound speed of the gravitino(s) \footnote{In a general theory of many interacting fields, the scalar sound speed $c_s$ may be understood as the determinant of the matrix of sound speeds of all fields kinetically coupled to the gravitino and with mass below the UV cutoff. } must be non-vanishing, $c_{s} > 0$.}
\end{quote}
This conjecture, analogous to the `No Global Symmetries' one of quantum gravity \cite{Kallosh:1995hi,ABBOTT1989687,GIDDINGS1988890, Banks:2010zn}, is easy to satisfy but surprisingly constraining.  In this {\it Letter} we discuss the evidence and implications of the above conjecture.\\
 
{\bf Catastrophic Production of Slow Gravitinos.}
A massive spin-3/2 field can be composed into propagating helicity-$1/2$ and helicity-$3/2$ components. The sound speed of helicity-1/2 gravitino's is given by \cite{Kolb:2021xfn}
\begin{align}\label{eq:r2again}
c_s^2 = \frac{(p-3m^2\Mpl^2)^2}{(\rho+3m^2\Mpl^2)^2} + \frac{ 4 \Mpl^4  \dot{m}^2}{(\rho+3m^2\Mpl^2)^2} \com
\end{align}
where $p$ and $\rho$ are the pressure and energy density of the field content of the theory, defined by the diagonal components of the stress tensor $\tensor{T}{^\mu_\nu}$, and dot denotes the derivative with respect to time. This applies in full generality, including in supergravity and in the Rarita-Schwinger model, and is in agreement with results in the supergravity literature \cite{Giudice:1999yt,Giudice:1999am,Kallosh:1999jj,Kahn:2015mla}.  The reduced sound speed, $c_s ^2 \leq 1$, follows from the canonical kinetic term for the gravitino after imposing the constraint equations. Unlike the EFT of inflation \cite{Cheung:2007st} and $P(X)$  theories \cite{ArmendarizPicon:2000dh,ArmendarizPicon:2000ah}, the sound speed is not generated by self-interactions or by irrelevant operators.  Thus, in contrast with those cases, the vanishing sound speed for the gravitino does not imply strong coupling or a breakdown of perturbative unitarity. Finally, we note that gravitino sound speed is distinguished from that in other theories by the dependence of $c_s$ on the mass and cosmological parameters, which leads to a violation of adiabaticity and hence particle production whenever the sound speed vanishes  \cite{Kolb:2021xfn}.

In a cosmological context, the evolution of the sound speed is dictated by the evolution of the background cosmology. For example, for the Rarita-Schwinger model in an FRW universe with equation of state $w\equiv p/\rho$, the sound speed is given by,
\begin{equation}
 c_s ^2 = \left( \frac{m^2 - w H^2}{m^2 + H^2} \right)^2 .
 \end{equation}
 In a radiation dominated universe, $w=1/3$, and with $m < H_e$ at an initial time $t_i$, after inflation, the redshifting of the radiation causes the sound speed to vanish at a time $t_* > t_i$ defined by $H(t_*) = \sqrt{3}m$.
 
The cosmological dynamics of the gravitino can be studied in a precise manner within the context of inflationary cosmology. In this case one may impose Bunch-Davies initial conditions on field excitations deep inside the horizon during inflation, and evolve the field forward in time to the end of inflation. The end stage of inflation is characterized by a period of oscillations of the inflaton field, which has been extensively studied in the context of preheating \cite{Traschen:1990sw,Kofman:1994rk,Kofman:1997yn}. The energy density and pressure during this phase are given by $\rho = \half\dot{\phi}^2 + \half m^2_\phi\phi^2$ and $p = \half\dot{\phi}^2 - \half m^2_\phi\phi^2$, where we have assumed for simplicity that the minimum of the potential is locally quadratic. The inflaton mass is related to the expansion rate at the end of inflation, $H_e$, by $m_\phi = 2H_e\Mpl/\phi_e$. The sound speed for the massive Rarita-Schwinger field is given by the first term of Eq.~\eqref{eq:r2again}.  If $m \lesssim H_e$, the sound speed vanishes once every oscillation. 
 
One might wonder if the vanishing of the sound speed is particular to the Rarita-Schwinger model, and is ameliorated by supergravity.  In an ${\cal N}=1$, $d=4$, supergravity model with $N$ chiral superfields, with evolving radial scalar components, the sound speed is given by \cite{Kolb:2021xfn},
\begin{align}
c_s^2 = 1 -  \frac{ 4 |\vec{\dot{\Phi}}|^2 |\vec{F}|^2  }{(|\vec{\dot{\Phi}}|^2 +|\vec{F}|^2)^2} \left( 1- \cos^2 (\theta) \right) \com
\end{align}
where $\theta$ is the angle between the field-space vector of the cosmological evolution, $\vec{\dot{\Phi}}$, and the $F$-term vector $\vec{F}$, and he norm of these vectors is taken with respect to the field space metric $G_{I \bar{J}}$. The sound speed vanishes when the direction of SUSY breaking is orthogonal to the direction of cosmological evolution and the two ingredients contribute equally to the energy density of universe.

An interesting example is the class of supergravity models containing a constrained nilpotent superfield, ${\bf S}(x,\theta)$, satisfying ${\bf S}^2(x,\theta)=0$. This class of models is the low energy EFT of an anti-D3 brane \cite{Kallosh:2014wsa,Bergshoeff:2015jxa,Vercnocke:2016fbt,GarciadelMoral:2017vnz, Cribiori:2019hod}, as famously utilized in the KKLT construction of dS vacua in string theory \cite{Kachru:2003aw}. As an example, consider a model specified by a K\"{a}hler potential  $K$ and superpotential $W$ given by \cite{McDonough:2016der,Kallosh:2017wnt}
\begin{equation}
\label{eq:nilpotentkahler}
W = M S+W_0 \Mpl \com  K = \Mpl^2 \frac{S \bar{S}}{f(\Phi,\bar{\Phi})} - 3 \alpha \log \frac{ \Phi + \bar{\Phi}}{\sqrt{4\Phi \bar{\Phi}}}
\end{equation}
where $S$ and $\Phi$ are chiral superfields, $f$ is a real function of $\Phi$ and $\bar{\Phi}$, and $W_0$, $M$, are constants. This model generically leads \cite{McDonough:2016der,Kallosh:2017wnt} to $\alpha$-attractor inflation \cite{Ferrara:2013rsa,Kallosh:2013yoa,Kallosh:2014rga} in the $\phi \equiv \sqrt{2} |\Phi|$ direction, after which $\phi$ oscillates about a point $\phi_*$ determined by the form of $f(\Phi,\bar{\Phi})$. The scalar potential is given by $V_{\rm tot} = \Lambda + V(\phi)$, where $\Lambda =M^2 - 3 W_0 ^2$ is the cosmological constant and $V(\phi)=M^2f/\Mpl^2$ is the inflaton potential. The inflatino mass $m_{\rm inflatino} = 3m\alpha (\Mpl/\phi)^2/2$ is, after inflation, set by $\phi_*$ and for simplicity may be taken to be heavy. The gravitino mass, $m=W_0/\Mpl$, is ostensibly sequestered into $\Lambda$, and thus naively independent of the physics of inflation. However, at the end of inflation the oscillations of $|\vec{\dot{\Phi}}|$ have amplitude of approximately $\Mpl H_e$, while the SUSY breaking has amplitude  $|\vec{F}| =M $, and the field space evolution and SUSY breaking are orthogonal at all times, $\vec{\dot{\Phi}} \cdot \vec{F}=0$.  If $M  \lesssim \Mpl H_e$, then the sound speed $c_ s^2$ vanishes once every oscillation, when $|\vec{\dot{\Phi}}|=|\vec{F}| $. Meanwhile, the observed near-vanishing of the cosmological constant, $\Lambda\simeq0$, dictates that  $M \simeq \sqrt{3} m \Mpl$.  From this one finds that for $m \lesssim H_e$ the sound speed vanishes in the nilpotent superfield model Eqs.~\eqref{eq:nilpotentkahler}.

\begin{figure*}
\label{fig:fig1}
\includegraphics[width=0.49\textwidth]{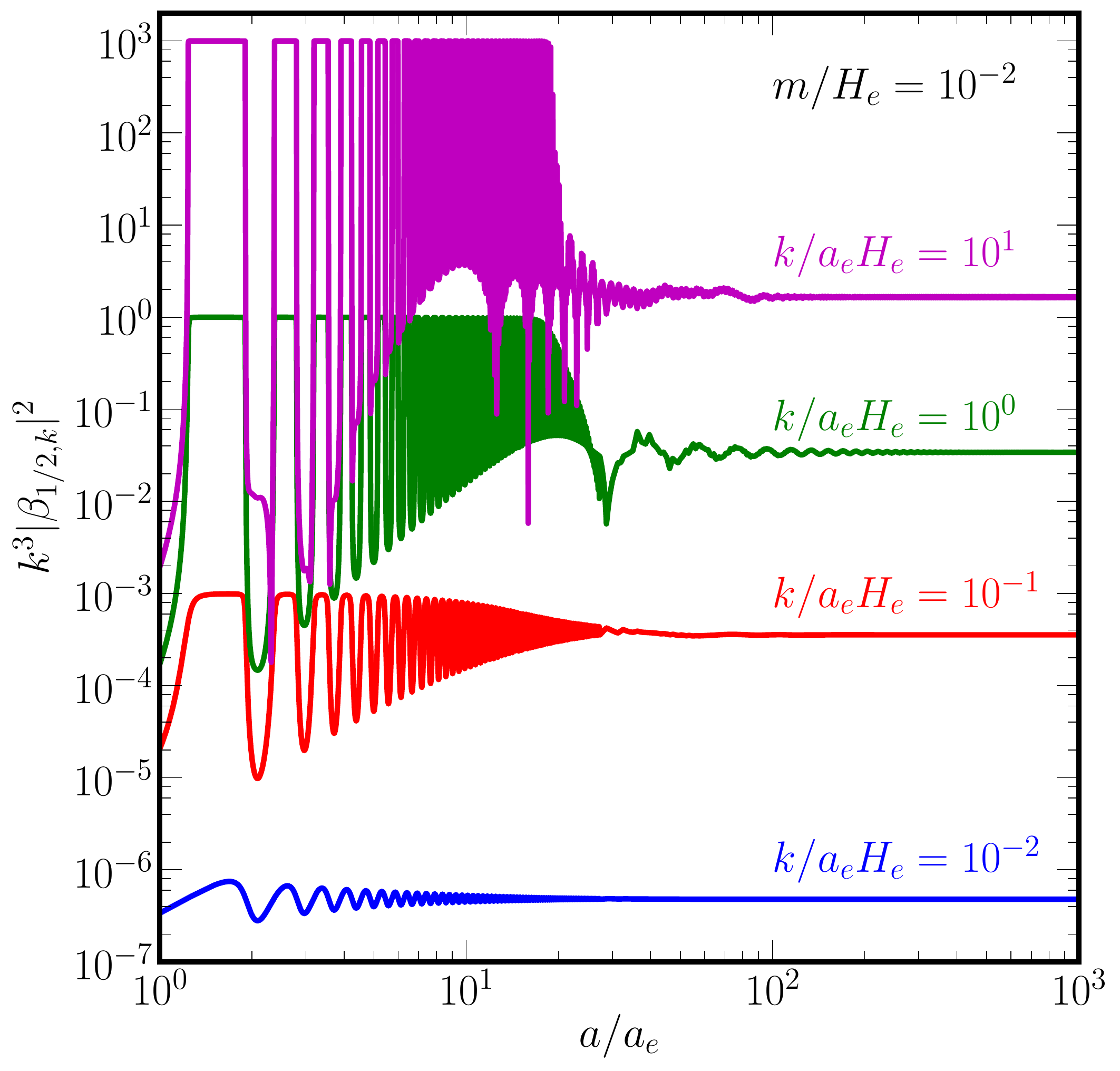}
\includegraphics[width=0.49\textwidth]{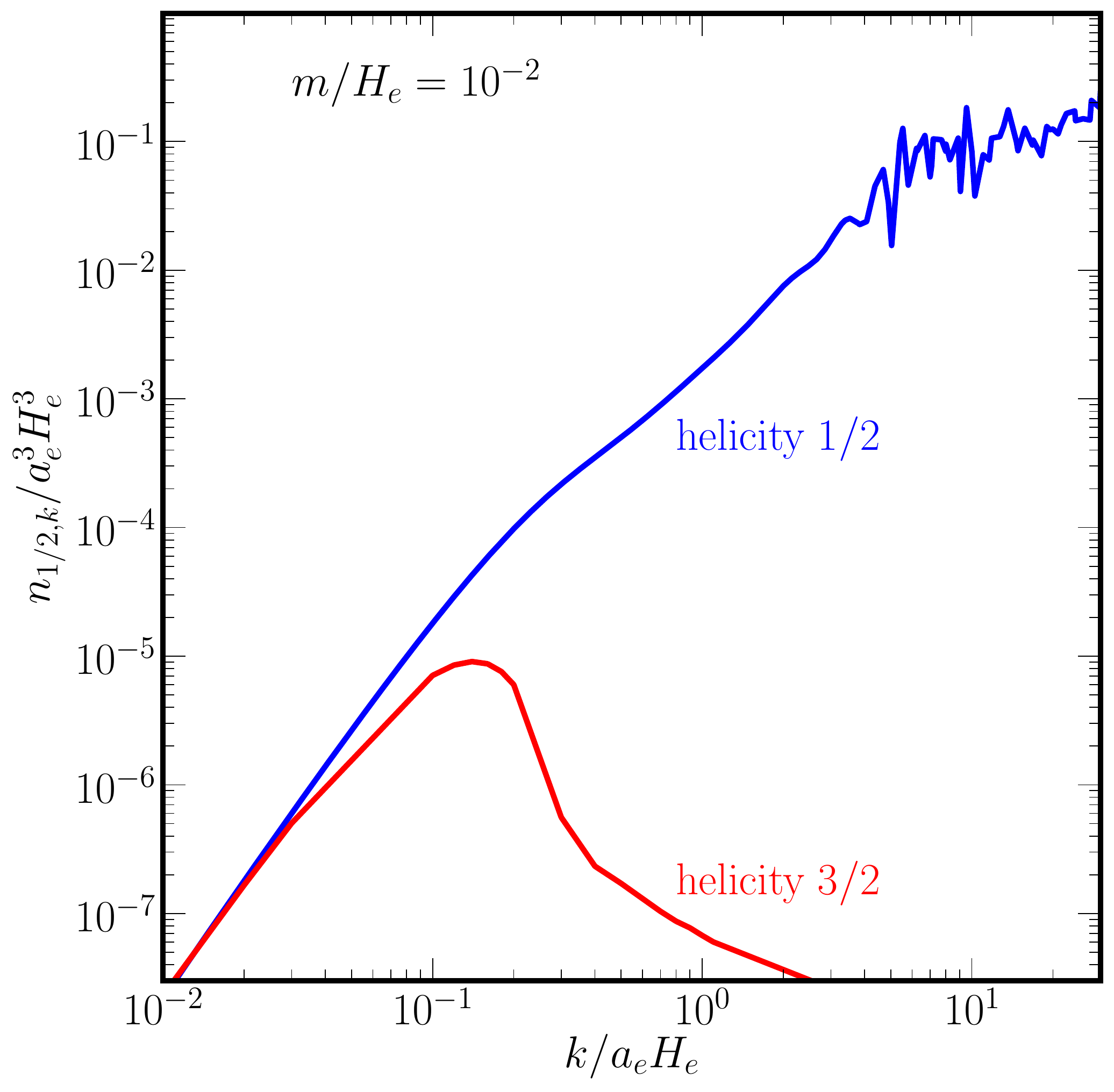}
\caption{The production of helicity-1/2 gravitinos in the post-inflationary universe for $m/H_e=10^{-2}$. {\it Left:} The evolution of $k^3 |\beta_{\Half,k}|^2$ as a function of the scale factor, for differing values of $k$.  {\it Right:} The spectrum of particles, defined by the late-time limit of $k^3 |\beta_{\Half,k}|^2$.  }
\end{figure*}
 
The vanishing of the sound speed leads to catastrophic gravitational particle production. This can be seen numerically by solving the mode equations of the gravitino, and understood analytically.  The spectrum of helicity-1/2 particles that results is calculated from the late-time amplitude of the negative-frequency modes as~\cite{Chung:2011ck}
\begin{equation}\label{eq:nk_12}
	n_{\Half,k} = \frac{k^3}{2\pi^2} \, |\beta_{\Half,k}|^2 ,
\end{equation}
for $|\beta_{\Half,k}|^2 \equiv \lim_{\eta \to \infty} |\chi_{-,\Half,k}(\eta)|^2 $, with $\chi_{-,\Half,k}(\eta)$ the negative frequency mode function. Here $n_{\Half,k}$ is the comoving number density of helicity-1/2 particles per logarithmically-spaced wavenumber interval.  

A numerical solution for a fiducial example is shown in Fig.~\ref{fig:fig1} for $m/H_e = 0.01$.  The spectrum of particles is shown in the right panel, where one may appreciate that $n_{\Half,k}$ increases without bound as $k$ increases. At high $k$, deep inside the horizon, the spectrum also exhibits high-frequency oscillations, which are familiar from past studies of gravitational production of gravitinos, see e.g., Giudice, Riotto, and Tkachev \cite{Giudice:1999am}.

To understand the time-evolution of this process,  the left panel of Fig.~\ref{fig:fig1} shows the time evolution of $k^3 |\beta_{\Half,k}|^2$ for differing values of $k$, which in the late-time limit determines the particle number $n_k$. One may appreciate that the dramatic rise of $k^3 |\beta_{\Half,k}|^2$ is synchronized across $k$-modes. This strongly suggests that the production of high-$k$ modes is not obstructed by backreaction of the produced particles on the inflaton oscillations, since there is no intermediary time window between the production of low-$k$ modes and the production of high-$k$ modes wherein backreaction of the former could shut off the production of the latter.

Gravitational particle production with vanishing sound speed may be understood analytically in terms of the violation of adiabaticity.   If adiabaticity is violated for a transient period of time, the generic outcome at later times once adiabaticity is restored is the presence of particles with respect to the late-time vacuum. For detailed studies see, e.g., Refs.~\cite{Amin:2015ftc, Garcia:2019icv, Garcia:2020mwi}. For illustration we consider a scalar field with dispersion relation $\omega_k ^2 = c_s ^2 k^2 + a^2 m^2$. The condition for adiabatic evolution of the vacuum state of the theory is given by $A_k \equiv |\dot{\omega}_k | / \omega_k ^2 \ll 1 $. At the moment when $c_s=0$, the adiabaticity condition is independent of $k$, and is simply $H/m$, so adiabaticity is clearly violated for $m<H_e$.  This property is particular to the gravitino, arising due to the relationship between $m$, the vanishing of $c_s$, and $A_k (c_s)$. In other theories where the evolution of $c_s$ is decoupled from $m$, e.g., a scalar field theory of the form ${\cal L}= P(X) + V(\phi)$, no such connection exists. From this we infer that the production of arbitrarily high-$k$ modes, referred to by   \cite{Kolb:2021xfn}  as ``catastrophic production,'' is a special property of quantized gravitinos.

{\bf Breakdown of Effective Field Theory.}
The vanishing of the gravitino sound speed generates particles of arbitrarily high momentum $k$, including $k$ equal to the cutoff of the EFT, namely the UV cutoff of four dimensional supergravity, corresponding to the scale where quantum gravity effects become important. This necessitates the inclusion of new, heavy, degrees of freedom with mass at or above the cutoff. The necessity of including these new degrees of freedom signals the breakdown of the  EFT.
 
One may also understand this in terms of irrelevant operators in the effective field theory that cannot be ignored once there are on-shell degrees of freedom with momentum at the cutoff.  These operators are not present in action of ${\cal N}=1$ $d=4$ supergravity (see, e.g. \cite{Freedman:2012zz}) but may descend from the UV completion. Any such corrections to the theory must vanish in the limit $m\rightarrow 0$, since $m$ is the order parameter of the spin-3/2 gauge invariance, and thus the subleading operators start at dimension $>5$, and are irrelevant operators. For example, a set of operators labelled by $n$,
\begin{equation} \nonumber
{\cal O}_n \propto c_n m \frac{\partial^{n}\left( \bar{\psi}^\mu  \psi_\mu \right)}{\Lambda^{n}}   \sim c_n m  \int d^3k   \left( \frac{k}{\Lambda}\right)^n \bar{\psi} ^\mu _{\bf k} \psi_{\mu{\bf k}} ,
\end{equation}
where $c_n$ are a set of order-1 coefficients, and $\Lambda$ is the UV cutoff the EFT (i.e, the UV cutoff of four dimensional supergravity).  Clearly these operators contribute non-negligibly to scattering amplitudes once particles with momentum $k \sim \Lambda$ are produced, and once $k=\Lambda$ is produced, the infinite tower must be included. Thus the theory ceases to be effective --- referred to colloquially as the ``breakdown of effective field theory.''

This stands in stark contrast to the conventional gravitino problem \cite{KHLOPOV1984265,ELLIS1984181}, wherein the EFT is valid at all times. We additionally note that the breakdown of the EFT is independent of the direction of supersymmetry breaking, and thus the particular (in principle time-dependent) combination of the spin-1/2 fermions that comprise the helicity-1/2 gravitino in the unitary gauge.

{\bf The Gravitino Swampland Conjecture (GSC).} In light of the breakdown of EFT that occurs when the gravitino sound speed vanishes, in the Introduction we proposed the Gravitino Swampland Conjecture.  The GSC holds that the gravitino sound speed cannot vanish in EFTs that have a UV completion in a theory of quantum supergravity, such as string theory. Despite being seemingly simple to evade due its nature as a non-vanishing condition, our analyses demonstrate the GSC has remarkable power: it relegates to the swamp massive Rarita-Schwinger models and the constrained superfield models with a heavy inflatino and a constant gravitino mass smaller than the expansion rate at the end of inflation.

The GSC is supported by the KKLT  \cite{Kachru:2003aw}  and Large Volume (LVS) \cite{Balasubramanian:2005zx}  scenarios  for moduli stabilization in string theory. In KKLT, stabilization of the radial modulus requires $m\gtrsim H$ at all times, while the LVS requires an even stronger constraint, $m \gtrsim (H^2 \Mpl)^{1/3}$.  
String theory satisfies the GSC in an additional way, through $\alpha'$ and $g_s$ corrections to the K\"{a}hler potential. As emphasized in \cite{Sethi:2017phn}, when supersymmetry is broken, perturbative corrections to the K\"{a}hler potential generate a scalar potential. They also generate a time-dependence of the gravitino mass   \cite{Kolb:2021xfn}. From Eq.~\eqref{eq:r2again}, the induced time-dependence of the gravitino mass lifts the zeros of $c_s ^2$ for any choice of $m/H_e$.

Thus, it certainly appears to be the case that $c_s ^2$ is non-vanishing in any effective field theory descended from string theory, and thus the GSC holds true. The next question is whether string theory provides a lower bound on $c_s ^2$, e.g., $c_s ^2 \gtrsim 1/\alpha$ with $\alpha$ an order-1 number, analogous the upper bound on axion decay constants from the Weak Gravity Conjecture \cite{ArkaniHamed:2006dz}.  One might reasonably call this the Strong Gravitino Swampland Conjecture. We leave this interesting possibility to future work.

{\bf Implications for Cosmology and Particle Physics.} 
Finally, we turn to the implications for cosmological observations and particle physics experiments. Particle colliders search for the gravitino, while cosmic microwave background experiments seek to measure the expansion rate during inflation, via the B-mode polarization generated by primordial gravitational waves \cite{Guzzetti:2016mkm}.  We take as the null hypothesis a constant mass gravitino, no light superpartners at the end of inflation that kinetically couple with the gravitino, and a conventional inflationary thermal history. Depending on the outcome of future experiments, one may draw striking conclusions:
 \begin{enumerate}
 \item If a gravitino is observed at a terrestrial particle collider, $m \lesssim {\rm TeV}$, then cosmological experiments will never observe the B-mode polarization of the cosmic microwave background generated by primordial gravitational waves.
 \item If B-mode polarization of primordial gravitational waves is observed, then collider experiments will never see a gravitino.
 \end{enumerate}
An exciting alternative possibility is that {\it both}   the gravitino and primordial B-modes are observed. In this case, we may reject the null hypothesis in favor of other possibilities, such as modifications to the thermal history of the universe (see e.g., \cite{Allahverdi:2020bys}). One possibility is for inflation to exit to a primordial black hole dominated phase, which reheats the universe much later, through Hawking radiation, while in the intervening time, when $w=0$, $H$ transits from $H \gg m$ to $H \ll m$ without $c_s ^2$ ever touching zero.

{\bf Discussion.} 
In this letter we have put forward a conjecture to delineate the boundary between effective supergravity theories that are low energy limits of quantum supergravity, and those that are not. We have proposed the Gravitino Swampland Conjecture, which states that the sound speed of gravitinos is non-vanishing. This is supported by gravitational production of gravitino quanta in cosmological spacetimes, and by examples in string theory. Applied to the massive Rarita Schwinger model, the GSC demands that the gravitino mass  satisfies $m \gtrsim H_e$, where $H_e$ is the Hubble expansion rate at the end of inflation. The GSC has surprisingly strong implications for the inferences we may draw from future discoveries at particle physics and cosmology experiments.

In the future it will be interesting to consider the interplay of the GSC with other swampland conjectures, such as the Swampland Distance Conjecture \cite{Ooguri:2006in}, the Weak Gravity Conjecture \cite{ArkaniHamed:2006dz}, and the de Sitter Swampland Conjecture \cite{Obied:2018sgi}.  Along these lines, in \cite{Lanza:2020qmt} it was shown that, in a specific context, the Weak Gravity Conjecture implies the Swampland Distance Conjecture. It would be interesting to see if something similar emerges for the gravitino.

The models which do satisfy the GSC may be phenomenologically interesting as a genesis mechanism for gravitino dark matter. We leave this interesting topic to future work. 

Finally, we note that alternative formulations of supergravity may exist that lead to a dispersion relation for the gravitino that is well behaved even when $c_s ^2$ (or in a coupled multi-fermion system, the determinant of the matrix of sound speeds) vanishes. Whether these formulations exist as a low-energy limit of string theory is an open and interesting question.

{\bf Acknowledgements.} The authors thank Mustafa Amin, Sylvester James Gates, Jr., Gian Giudice, Wayne Hu, Andrei Linde, Antonio Riotto,  Marco Scalisi, Leonardo Senatore,  Gary Shiu, Igor Tkachev, Cumrun Vafa, and Lian-Tao Wang for helpful comments.  The work of E.W.K.\ and E.M.\ was supported in part by the US Department of Energy contract DE-FG02-13ER41958.

\bibliography{swamp}
\bibliographystyle{JHEP}

\end{document}